# Role of Functionalized Graphene Quantum Dots in Hydrogen Evolution Reaction: A Density Functional Theory Study

Vaishali Sharma[1,*], Basant Roondhe[2], Sumit Saxena[2] and Alok Shukla[1,*]

[1]Department of Physics, Indian Institute of Technology Bombay, Powai, Mumbai 400076, India
[2]Nanostructures Engineering and Modeling Laboratory, Department of Metallurgical Engineering and Materials Science, Indian Institute of Technology Bombay, Mumbai, MH 400076, India

*Corresponding author email: oshivaishali@gmail.com, shukla@phy.iitb.ac.in

**Abstract:** Rapid advances in the field of catalysis require a microscopic understanding of the catalytic mechanisms. However, in recent times, experimental insights in this field have fallen short of expectations. Furthermore, experimental searches of novel catalytic materials are expensive and time-consuming, with no guarantees of success. As a result, density functional theory (DFT) can be quite advantageous in advancing this field because of the microscopic insights it provides and thus can guide experimental searches of novel catalysts. Several recent works have demonstrated that low-dimensional materials can be very efficient catalysts. Graphene quantum dots (GQDs) have gained much attention in past years due to their unique properties like low toxicity, chemical inertness, biocompatibility, crystallinity, etc. These properties of GQDs which are due to quantum confinement and edge effects facilitate their applications in various fields like sensing, photoelectronics, catalysis, and many more. Furthermore, the properties of GQDs can be enhanced by doping and functionalization. In order to understand the effects of functionalization by oxygen and boron based groups on the catalytic properties relevant to the hydrogen-evolution reaction (HER), we perform a systematic study of GQDs functionalized with the oxygen (O), borinic acid ($BC_2O$), and boronic acid ($BCO_2$). All calculations that included geometry optimization, electronic and adsorption mechanism, were carried out using the Gaussian16 package, employing the hybrid functional B3LYP, and the basis set 6-31G(d,p). With the variation in functionalization groups in GQDs, we observe significant changes in their electronic properties. The adsorption energy $E_{ads}$ of hydrogen over O-GQD, $BC_2O$-GQD, and $BCO_2$-GQD is -0.059 eV, -0.031 eV and -0.032 eV respectively. Accordingly, Gibbs free energy (ΔG) of hydrogen adsorption is extraordinarily near the ideal value (0 eV) for all the three types of functionalized GQDs. Thus, the present work suggests pathways for experimental realization of low-cost and multifunctional GQDs based catalysts for clean and renewable hydrogen energy production.

# Introduction

Continuous increase in the consumption of fossil fuels to fulfill ever-increasing energy needs of humanity has led to enormous pollution and degradation of the environment. This has prompted a worldwide effort by researchers to explore novel clean and renewable energy sources. The research has suggested hydrogen to become the first-choice fuel, with the potential to replace fossil fuels, due to its abundance in nature, high energy density, and most importantly its environmental friendliness [1-2]. Several methods have been developed for producing hydrogen of which electrochemical water splitting has demonstrated great promise because of its high hydrogen production efficiency, purity and above all absence of other greenhouse gases or pollutant production in the process [3]. The half reactions for water-splitting, the hydrogen evolution reaction (HER) is difficult to conduct unless the process is assisted by an active catalyst to effectively minimize the required overpotential which is the main concern. From past decades, electrocatalysis studies have received active attention by both experimental as well as computational groups [3] because of their possible applications in developing renewable and clean energy through energy harvesting. The economic feasibility of electrocatalytic water splitting is crucial and can be achieved by decreasing both the energetic cost (i.e., overpotential) and the material cost for the two half-reactions, the oxygen evolution reaction (OER) and the hydrogen evolution reaction (HER) [4-6]. Nano-sized materials can be quite efficient as catalysts due to their significant advantages such as low fabrication cost and high specific surface area to reach the surface reaction area. Generally, Pt-based materials have been regarded as the most effective catalysts for HER [7-8], however, they cannot be used widely and practically due to their scarcity and high cost. Therefore, there is an urgent need to develop inexpensive alternatives with high HER catalytic activity. To date, some relatively inexpensive HER catalysts have been investigated [9–22] such as transition-metal sulfides and phosphides [9-10], pnictides [11], borides [12], carbides [13-14], nitrides [15-16], as well as carbon materials [17-20], and metal–organic complexes [21-22].

Carbon quantum dot (CQDs) have attracted tremendous attention in the past decade due to their potential in bio-imaging [23-24], photodetectors [25], photocatalysis [26-28] and energy related applications [29]. Generally, CQDs including carbon nanodots [30] and graphene QDs (GQDs) [31] have multifold advantages, such as facile synthesis and excellent chemical inertness. Most importantly, GQDs are graphene fragments with desirable optical and electronic properties. Graphene has shown high electron mobility along with superior mechanical and thermal stability making it suitable for many applications, [32-34] but the key deficiency in graphene is its zero band gap which limits its applications in photonics and optoelectronics. Scientists have proposed many approaches by which a band gap can be opened in graphene. One of the effective and largely adopted methods is by reducing the dimensions of graphene into GQDs. The band gap of GQDs can be made to vary from 0-6 eV by: (a) adjusting their shapes and sizes, and (b) by attaching suitable functional groups or ad-atoms [26]. The edge chemistry, surface, and size of GQDs can be effortlessly changed in a very wide range by utilizing different synthesis approaches and subsequent treatments. Among all methods used to the tailor electronic properties of GQDs, heteroatom doping is a very effective strategy to alter the electronic properties of nanomaterials [35]. Recently, GQDs with transition metal nanoparticles adsorbed on their surface have been used in hydrolysis for effective hydrogen evolution [36]. Introduction of various heteroatoms results in enhanced catalytic activity due to the fact that they can intensely alter the local spin or charge distribution by carrying an uneven charge distribution in the material [37]. There are several reports on various types of heteroatom doped GQDs like nitrogen (N) doping which is broadly used to tune the typical properties of GQDs as the nitrogen atom has five valence electrons with a similar atomic radius to the carbon atom. Apart from N, the boron (B) atom also has a similar atomic radius which makes it relatively easy to bond with carbon leading to negligible structural disorder. As a result, B doping of GQDs has attracted significant research interest for its novel properties [38-43]. Additionally, the fabrication method depends on the use of strong oxidizing agents which results in the unavoidable introduction of oxygen-containing functional groups attached to GQDs. These mainly include carboxyl (−COOH), epoxy (−O−), carbonyl (−(CO)−), ether (−OCH$_3$), and hydroxyl (−OH) groups [44]. These attached oxygen containing functional groups significantly enhance the solubility and also alter the electronic and optical properties of GQDs. Thus, making it interesting to see its effect on the HER activity of GQDs. Nevertheless, there is a lack of understanding of the electronic and HER properties of boron and oxygen functionalized GQDs. In this work, we aim to probe the influence of functionalization on the HER activities of GQDs. However, there is a lack of understanding in how the electronic and HER properties of boron and oxygen functionalized GQDs change. In this work, we undertake systematic first-principles DFT to study the effects of three types of chemical configurations containing B and O atoms in the form of borinic acid (BC$_2$O) and boronic acid (BCO$_2$) and oxygen (O) functional groups, on the electronic and catalytic properties of GQDs. Our findings will facilitate further and comprehensive exploration of the HER mechanism of B-doped (BC$_2$O and BCO$_2$) and O-doped GQDs as a function of doping configurations, thereby providing insights into the design of more efficient GQD based catalytic devices.

## Computational Methodology

The geometry optimization and electronic properties calculations were performed using the density functional theory (DFT) as implemented in the Gaussian16 program package [45], along with the B3LYP hybrid exchange-correlation functional. The Gaussian-type split valence basis set 6-31G (p,d), which includes the polarization functions, was employed in the calculations. For geometry optimization, all considered systems were allowed to relax simultaneously until the gradient forces reached the set threshold value of 0.000450 Hartree/Bohr, RMS force of 0.000300 Hartree/Bohr, maximum displacement of 0.001800 Bohr, and RMS displacement of 0.001200 Bohr. The self-consistent field (SCF) convergence is set to $10^{-8}$ Hartree. The vibrational frequencies were also evaluated to confirm the absence of imaginary frequencies, in order to ensure the stability of the optimized geometries. Molecular orbitals (MO's) such as highest occupied molecular orbitals (HOMO) and lowest unoccupied molecular orbitals (LUMO) were visualized using with GaussView [46]. The Multiwfn software was employed to generate projected density of states (PDOS) [47]. The adsorption energy was computed using the formula:

$$E_{ads} = E_{system+H_2} - (E_{system} + E_{H_2}) \qquad (1)$$

Where $E_{system+H_2}$ denotes total energy of hydrogen molecule over O-GQD/BC$_2$O-GQD/BCO$_2$-GQD systems, $E_{system}$ represents optimized total energy O-GQD/BC$_2$O-GQD/BCO$_2$-GQD systems and $E_{H_2}$ represents optimized total energy of hydrogen molecule. A negative value of $E_{ads}$ implies a stable adsorption complex on the O-GQD/BC$_2$O-GQD/BCO$_2$-GQD systems. Additionally, in order to be sure about the final geometries, we performed one more iteration of geometry optimization (single point calculation) on the system with an adsorbed hydrogen molecule, at the same level of theory. The hydrogen evolution reaction (HER) is an electrochemical reaction consisting of proton reduction at the electrode to initially yield atomic hydrogen, resulting finally in the formation of the H$_2$ gas. The catalytic surface potential of the considered HER is calculated using the exchange current density linked to the free energy of adsorbed hydrogen (H$_2$) when the reaction is at equilibrium. The free energy of hydrogen in the adsorbed state can be computed using equation [48]:

$$\Delta G_{H^*} = E_{ads} + \Delta E_{ZPE} - T\Delta S \tag{2}$$

Above, $\Delta E_{ZPE}$ denotes zero-point energy (ZPE) corrections (ranging 0.01 eV to 0.04 eV for hydrogen molecule), $\Delta S$ denotes entropy difference and $T$ is the temperature. Further, the $\Delta E_{ZPE}$ can be written as [49]

$$\Delta E_{ZPE} = E_{ZPE}^{nH} - E_{ZPE}^{(n-1)H} - \frac{1}{2} E_{ZPE}^{H_2} \tag{3}$$

Here, $E_{ZPE}^{nH}$ and $E_{ZPE}^{(n-1)H}$ denote zero-point energy corrections of total energy of the systems with $n$ and $(n-1)$ hydrogen atoms adsorbed on the catalyst, respectively. $E_{ZPE}^{H_2}$ represents zero-point energy of gas phase $H_2$. $\Delta S$ in eq. (2) is the entropy variation among gas-phase hydrogen and adsorbed hydrogen which can be evaluated from the $H_2$ gas entropy [50] using the approximation

$$\Delta S = \left(S_{nH} - S_{(n-1)H} - \frac{1}{2} S_{H_2}\right) \approx \frac{1}{2} S_H^0 \tag{4}$$

Here, $S_{nH}$ is the entropy of the systems with $n$ hydrogen adsorbed and $S_{(n-1)H}$ is the entropy with $n-1$ hydrogen atoms adsorbed. $\frac{1}{2} S_{H_2}$ is the gaseous phase entropy of hydrogen. $S_H^0$ is the vibrational entropy of the hydrogen molecule which finally leads to the value of $TS_H^0 = 0.41$ eV [48]. Thus, we finally obtain the working formula:

$$\Delta G_{H^*} = E_{ads} + 0.24 \ eV \tag{5}$$

In order to develop high-quality catalyst for HER, the ideal value of $\Delta G$ should be nearly zero because a small $|\Delta G|$ shows superior HER efficacy. $\Delta G$ is one of the most significant parameters concerning HER performance of an electrocatalyst. The theory that ideal HER activity of an electrocatalyst should have a $\Delta G$ value of around zero was firstly given by Parson and later proved by others [51-54]. If the catalysts bind weakly, it will be problematic for the surface to activate them. However, if the catalysts bind too strongly, it may impede desorption of $H_2$ molecules by occupying all accessible surface sites, thus making the reaction ineffective. Therefore, for hydrogen evolution reaction, the optimum value of ΔG should be around zero. Additionally, the other important parameter relevant to HER is the value of $E_{ads}$, which should be in the vicinity of 0.24 eV. The overpotential (η) can be determined utilizing $\Delta G$:

$$\eta = -\frac{|\Delta G|}{e} \qquad (6)$$

**Results and Discussion**

The strong motivation for the present study arises from the previous studies of oxygen and boron functionalized GQDs [55-56]. In this work, we have considered the functionalization of the GQDs by boron and oxygen by doping them with borinic and boronic acids both of which have B–O bonds, and by the oxygen containing group (O) [57-58]. The three boron- and oxygen-doped geometries considered in this work are shown in Fig. 1. For convenience, oxygen, borinic acid, and boronic acid functionalized GQDs are expressed as O-GQD, $BC_2O$-GQD and $BCO_2$-GQD, respectively. Initially, the geometries of O-GQD, $BC_2O$-GQD and $BCO_2$-GQD are optimized individually. The optimized structures are presented in Fig. 1. After the optimization, near the attached functional groups, the C-C bond length changes from its standard value 1.42 Å to 1.5 Å in O-GQD, to ~1.48 Å in $BCO_2$-GQD, and 1.52 Å in $BC_2O$-GQD, in agreement with previous studies on GQDs [59-63]. The change in C-C bond length strongly depends on the atomic radius of the dopant. For instance, the atoms present in the functional groups considered here (oxygen and boron) have similar bond lengths and atomic radii as carbon, therefore they will attach with the carbon atoms with minimum deformation (with less change in C-C bond length). However, for larger atoms with large atomic radii, it forms $sp^3$-like bonding with C-atoms with noticeable deformation. The B-C bond lengths post-optimization are 1.52 Å, and 1.64 Å for $BC_2O$-GQD and $BCO_2$-GQD, respectively, again in good agreement with the earlier work [64]. As seen in Fig. 1, the functionalization in GQDs slightly distorts its hexagonal ring near O, $BC_2O$ and $BCO_2$ group which can be due to the inclusion of groups resulting in disruption of sp$^2$ hybridization transforming to sp$^3$ hybridization of carbon atoms in GQDs.

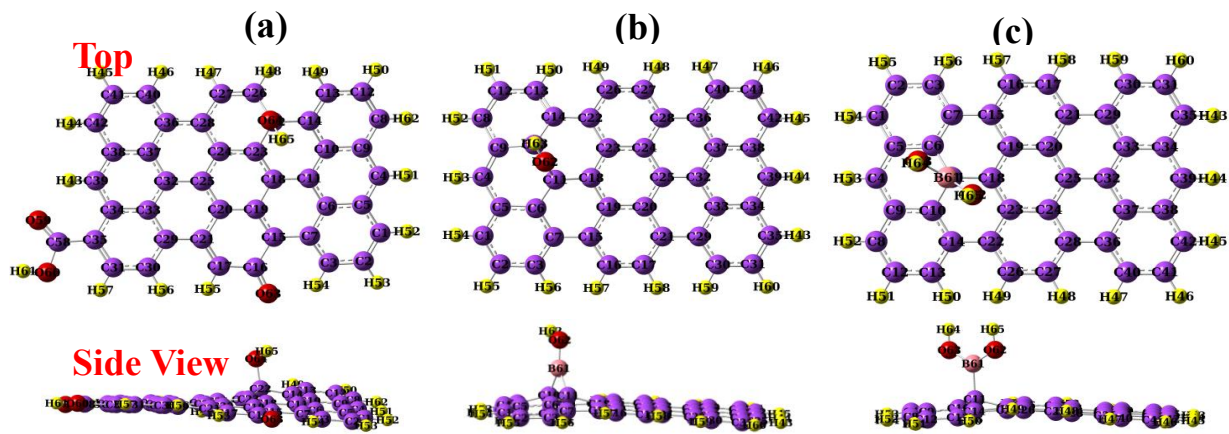

**Figure 1:** Optimized structures of (a) O-GQD, (b) BC$_2$O-GQD and (c) BCO$_2$-GQD. Yellow, purple, red and pink represent hydrogen, carbon, oxygen and boron atoms, respectively.

This results in a slight pyramid-like structure, i. e. buckling, above the planer structure of the GQDs, again in agreement with the previous studies [65-67]. Next, we re-optimized the geometries with an H$_2$ molecule adsorbed on top of each functionalized GQD. For the sake of completeness, in Figure 2 we present the initial geometries of O-GQD, BC$_2$O-GQD and BCO$_2$-GQD with a hydrogen (H$_2$) molecule on top of each structure, before the starting of geometry optimization iterations. In all three cases, the H$_2$ is placed near the functional groups above the surface, at a distance of 2 Å.

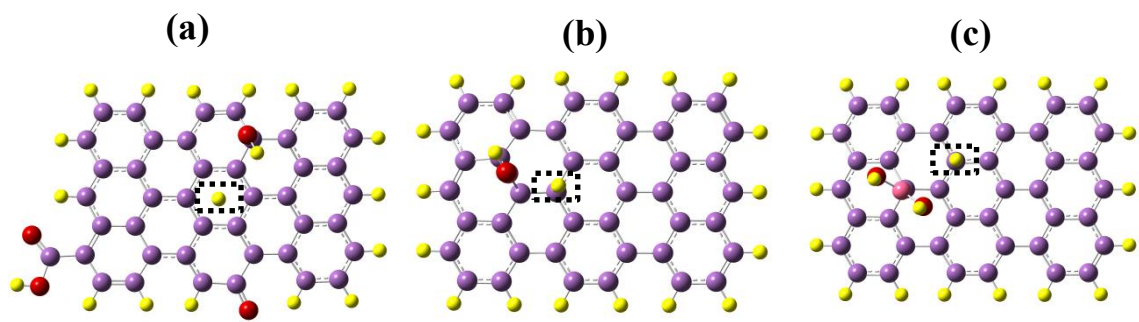

**Figure 2:** Initial structures of hydrogen over (a) O-GQD, (b) BC$_2$O-GQD and (c) BCO$_2$-GQD. For each GQD, the adsorbed hydrogen molecule is enclosed inside a dashed rectangular box.

Figure 3 shows the optimized structure of $H_2$ over O-GQD, $BC_2O$-GQD and $BCO_2$-GQD. We also performed vibrational frequency analysis on the final geometries, and the fact that no imaginary frequencies were observed implies that the optimized structures are stable and represent true minima. Following optimization, for $H_2$ over O-GQD ($H_2$-O-GQD), as seen from Fig. 3(a), the hydrogen atom is attracted towards the OH group attached to the ring. The distance of lower hydrogen atom (of hydrogen molecule) with respect to the adjacent carbon changes from 2 Å to 3.7 Å.

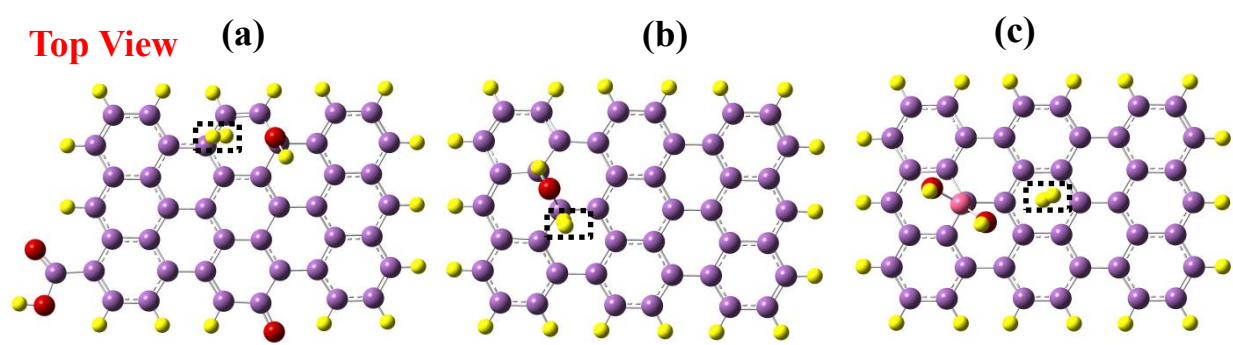

**Figure 3:** Optimized structures of hydrogen over (a) O-GQD, (b) $BC_2O$-GQD and (c) $BCO_2$-GQD. For each GQD, the adsorbed hydrogen molecule is enclosed inside a dashed rectangular box.

However, the distance between hydrogen and oxygen atom of functional group alters from 2.8 Å to 2.4 Å which is due to the highly electronegative nature of oxygen atom that tends to attract positive charges over hydrogen. For hydrogen over $BC_2O$-GQD ($H_2$-$BC_2O$-GQD), similar to $H_2$-O-GQD, the hydrogen molecule is attracted towards the borinic group, however, $H_2$ is relocated slightly upward as the distance between hydrogen and oxygen alters to 2.6 Å from 2.3 Å. In case of hydrogen over $BCO_2$-GQD ($H_2$-$BCO_2$-GQD), the $H_2$ is repelled from its original position and reoriented in the middle of benzene ring of $BCO_2$-GQD with C-H (carbon of benzene ring and hydrogen molecule) distance changing to 4.13 Å from 2 Å. The distance between oxygen (oxygen atom of functional group attached) and hydrogen adsorbed slightly increases (from 2.5 Å to 2.6 Å) after optimization. In the adsorbed $H_2$ molecule, H-H bond distance changes from 0.742 Å to 0.745 Å in the case of $H_2$-O-GQD, however, in cases of $H_2$-$BC_2$O-GQD and $H_2$-$BCO_2$-GQD, it changes to 0.743 Å. Next, the electronic properties of functionalized GQDs considered in this work are analyzed, from the point-of-view of their possible applications as electrocatalysts in HER. The computed dipole moments ($p$) of various structures are presented in Table 1.

**Table 1:** Calculated HOMO ($E_{HOMO}$), LUMO ($E_{LUMO}$), energy gap ($E_g$), dipole moment, work function ($\phi$) and H-H bond lengths in all the considered systems.

| System | $E_{HOMO}$ (eV) | $E_{LUMO}$ (eV) | $E_g$ (eV) | Dipole moment p (D) | $\phi$ (eV) | $d_{H-H}$ (Å) |
|---|---|---|---|---|---|---|
| O-GQD | -4.696 | -3.512 | 1.184 | 3.738 | 4.104 | - |
| $H_2$-O-GQD | -4.716 | -3.523 | 1.192 | 3.625 | 4.120 | 0.745 |
| $BC_2$O-GQD | -4.114 | -2.653 | 1.460 | 2.346 | 3.384 | - |
| $H_2$-$BC_2$O-GQD | -4.125 | -2.665 | 1.459 | 2.235 | 3.395 | 0.743 |
| $BCO_2$-GQD | -4.107 | -2.422 | 1.684 | 4.052 | 3.264 | - |
| $H_2$-$BCO_2$-GQD | -4.114 | -2.429 | 1.684 | 3.949 | 3.272 | 0.743 |

Dipole moments not only give us information about the polarity of a given molecule, but also about its anisotropy, and reactivity as far as electrocatalysis is concerned. It is obvious from Table 1 that $p$ is reduced in all the three structures, i.e., O-GQD, $BC_2$O-GQD and $BCO_2$-GQD subsequent to hydrogen adsorption. The value of $p$ varies from ~2.2 D to 4 D indicating: (a) polar nature of bonding in these systems, and (b) structural and electronic anisotropies. The decrement in $p$ indicates that the adsorption of hydrogen molecule in O-GQD, $BC_2$O-GQD and $BCO_2$-GQD leads to the decreased polarity in these systems. Additionally, $H_2$ adsorption in O-GQD, $BC_2$O-GQD and $BCO_2$-GQD induces charge transfer, resulting in the modification of $p$. It is well known that lower values of $p$ lead to relative stability suggesting that the hydrogen adsorbed GQDs are structurally more stable as compared to the original functionalized GQDs considered here [68].

To further understand the electronic properties of the considered structures, the energies of their highest occupied molecular orbitals (HOMO) and lowest unoccupied molecular orbitals (LUMO) are presented in Table 1, while the orbitals are plotted in Fig. 4.

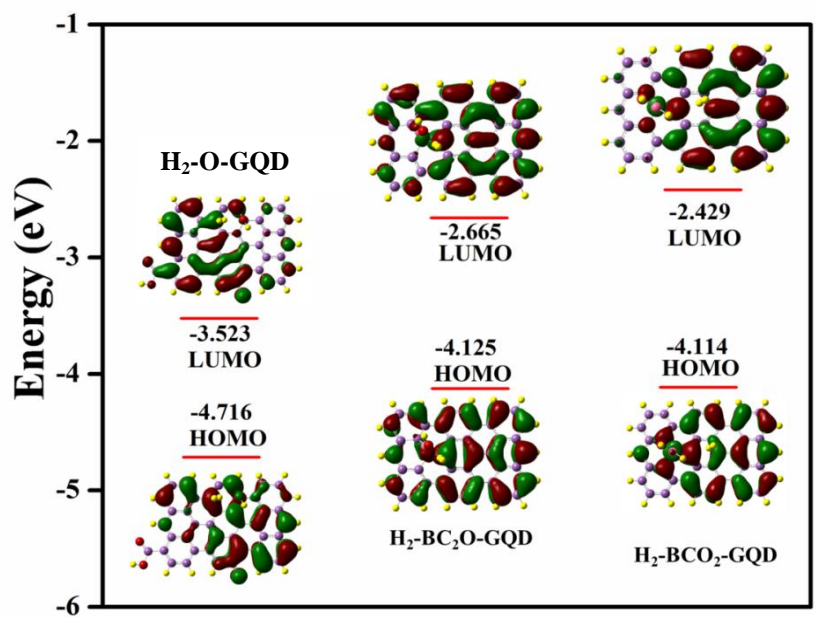

**Figure 4:** HOMO-LUMO plots of hydrogen over O-GQD, $BC_2O$-GQD and $BCO_2$-GQD.

The HOMO levels represent the electron donor properties of the system, and thus show oxidative stability or reactivity of electrophilic kind. The LUMO levels, on the other hand, denote the electron acceptor properties of the system, and, therefore, indicate the reactivity of the nucleophilic kind. In Table 1, we present the energies of the HOMO and LUMO levels along with the band gaps of all the systems considered in this work. From the table, it is obvious that hydrogen over O-GQD, $BC_2O$-GQD and $BCO_2$-GQD results in slightly reduced $E_{HOMO}$ values after adsorption depicting their reduced electron-donor ability. Similarly, $E_{LUMO}$ is also reduced depicting their reduced electron-acceptor ability. After the adsorption of the hydrogen molecule, extremely small to no change in the $E_g$ is observed indicating a moderate influence of $H_2$ adsorption on the band gaps of O-GQD, $BC_2O$-GQD and $BCO_2$-GQD. The values of the band gaps $E_g$ of $BC_2O$-GQD and $BCO_2$-GQD are in good agreement with the previously reported work [55]. An efficient electrocatalyst for HER should possess active sites for catalysis. As electrons contribute to HER process, the catalyst should have larger electronic conductivity which depends on the band gap ($E_g$) in a simple way: the larger the value of $E_g$, the lower the conductivity and vice versa. From Table 1, it is obvious that $H_2$-O-GQD has the smallest band gap as compared to the other two $H_2$ adsorbed GQDs. Therefore, comparatively, we expect it to have the highest electronic conductivity leading to the best electrocatalytic performance in HER [69-71]. Figure 5 presents total and projected density of states (PDOS) of O-GQD, $BC_2O$-GQD, and $BCO_2$-GQD, together with and without the adsorbed hydrogen molecule. The PDOS provides the understanding of occupied/unoccupied electronic levels, together with the spatial distribution of particular constituent atoms electronic orbitals. For O-GQD, it can be seen that carbon and oxygen mainly contribute to the HOMO levels, however, hydrogen contributes to LUMO energy levels. Similar behavior is visible in the case of $H_2$-O-GQD. In case of $BC_2O$-GQD, the contribution of carbon is solely to the HOMO level, while, hydrogen, boron and oxygen provide larger contributions to the LUMO level. Furthermore, in $H_2$-$BC_2O$-GQD, oxygen also makes significant contributions in the HOMO region. Additionally, similar PDOS curves are observed in case of $BCO_2$-GQD and $H_2$-$BCO_2$-GQD indicating major contributions of carbon, hydrogen, boron and oxygen atoms to the HOMO levels. It is obvious from the figure (Fig 5 (a-f)) that both TDOS and PDOS are higher in the HOMO region as compared to the LUMO region.

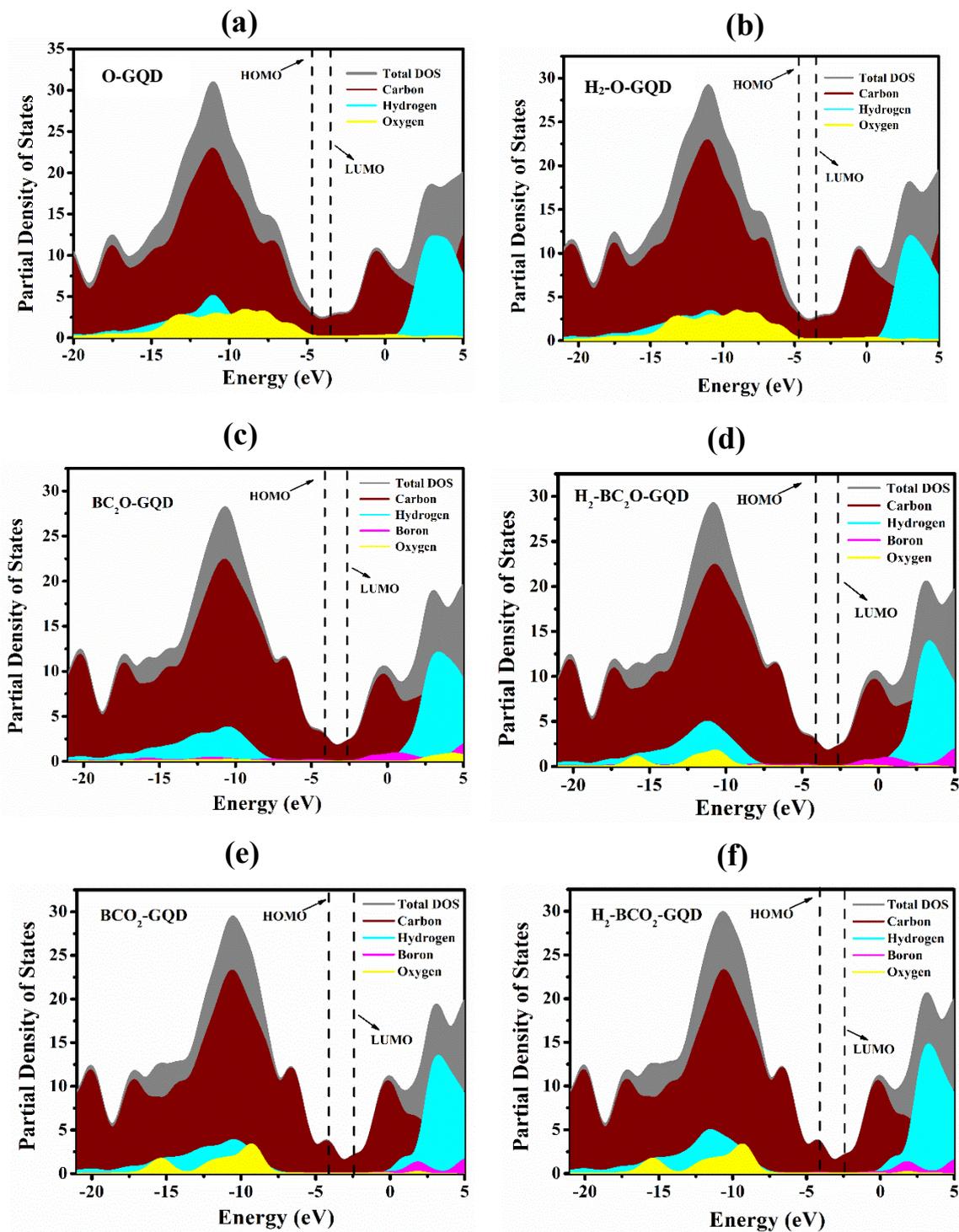

**Figure 5:** Total and atom projected density of states plots of (a) O-GQD, (b) BC$_2$O-GQD, (c) BCO$_2$-GQD (d) H$_2$-O-GQD, (e) H$_2$-BC$_2$O-GQD and (f) H$_2$-BCO$_2$-GQD.

It is also observed that the DOS plots in all the cases are very similar to each other from which it is evident that different functional groups and adsorbed hydrogen molecules have a minimal effect on the electronic structure of GQDs. The only changes one notices are small modifications in the positions of HOMO and LUMO energy levels. Thus, it can be said that the electronic structure of the GQDs is affected only near the Fermi level through the orbital hybridization caused by the functional groups, imparting the $sp^2$ hybridized carbon atoms with a bit of $sp^3$ character [55]. The work function ($\phi$) of a system is the minimum energy required to extract an electron from it to a point in a vacuum immediately outside the surface, without any kinetic energy [72]. Therefore, to attain an understanding of the influence of the adsorbed hydrogen on the reactive properties of O-GQD/BC$_2$O-GQD/BCO$_2$-GQD, $\phi$ has been calculated. For this purpose, the following equation was used:

$$\phi = \frac{Ionization\ potential + Electron\ affinity}{2} \qquad (7)$$

The ionization potential (IP) of a system can be estimated from its HOMO energy as IP = $-E_{HOMO}$, while the electron affinity can be estimated from its LUMO energy as EA= $-E_{LUMO}$ [73]. IP and EA values are significant in examining the electrochemical stability of the considered systems. The calculated values of $E_{HOMO}$, $E_{LUMO}$, and $\phi$ are tabulated in Table 1. It is found that $\phi$ of all the three structures increases by very small amounts after the hydrogen adsorption. For H$_2$-O-GQD, when compared to the other structures, we obtain the highest value of the work function 4.120 eV. This also suggests higher stability of H$_2$-O-GQD, as compared to the other two. Furthermore, our calculated value of $\phi$ for H$_2$-O-GQD is quite close to the work functions of Pt and Pd catalysts, suggesting that this system could prove to be an excellent catalyst for HER.

Having discussed the structural and electronic properties of the functionalized GQD systems, subsequently, their HER performances using the Sabatier principle are studied [74]. We consider the H$_2$ adsorption on O-GQD, BC$_2$O-GQD and BCO$_2$-GQD by placing the molecule near the functional groups to understand their contribution to the whole process. The interaction between $H_2$ and O-GQD, $BC_2O$-GQD and $BCO_2$-GQD is estimated by computing the adsorption energies $E_{ads}$ (eqn. (1)), and the results are presented in Table 2. $E_{ads}$ of $H_2$ is a significant parameter when analyzing the catalytic activity on the surface of the concerned catalyst. Among O-GQD, $BC_2O$-GQD and $BCO_2$-GQD, the highest $E_{ads}$ of -0.059 eV is obtained for $H_2$-O-GQD. Roughly half of that value, -0.03 eV, is found when it comes to $E_{ads}$ of $H_2$-$BC_2O$-GQD and $H_2$-$BCO_2$-GQD. For the efficient HER activity, it is important to produce a catalyst that does not attach strongly with the hydrogen in order to make the desorption step attainable, or, else, the objective of the hydrogen energy production will be impossible to fulfill. The calculated values of $E_{ads}$ suggest very feeble interaction of hydrogen with O-GQD, $BC_2O$-GQD and $BCO_2$-GQD, suggesting that these structures are good contestants for noble-metal free HER electrocatalysis. Generally, the higher $E_{ads}$ of $H_2$-O-GQD is the result of higher work function [75]. Next, using the values of work function, we compute the values of open-circuit potential V in an aqueous electrolyte using the equation:

$$V = \frac{\phi}{e} - 4.44 \qquad (8)$$

In above equation, 4.44 is the absolute potential value of the standard hydrogen electrode [76]. The open-circuit potential V values for all considered systems are tabulated in Table 2. The value ranges between -0.33 V to -1.16 V following the adsorption of hydrogen molecule. The water electrolysis device comprises anode and cathode as catalysts for HER and oxygen evolution reactions (OER) [77]. For the water-reduction reaction at the cathode, two mechanisms play important roles in the acidic media: (a) the Volmer−Heyrovsky, and (b) Volmer−Tafel mechanisms [77]. Regardless of the route acquired by HER, for both the mechanisms, $E_{ads}$ of hydrogen is an essential parameter defining the catalytic activity on the surface of catalyst. According to the Sabatier principle, the free energy (ΔG) is directly proportional to the exchange current density, which in turn is a measure of catalytic efficiency [43]. Because ΔG is connected to $E_{ads}$ through Eq. 5, both these quantities can be used to gauge the efficiency of the considered structures as catalysts for HER [78-79]. The values of ΔG of $H_2$-O-GQD, $H_2$-BC$_2$O-GQD and $H_2$-BCO$_2$-GQD, calculated using eqn. 5, are presented in Table 2. Along with that, Gibbs free energies of hydrogen evolution reaction are also calculated using thermochemistry analysis. The Gibbs free energies of reaction are calculated employing the formula [80]: $\Delta_r G°(298K) = \sum(\varepsilon_0 + G_{corr})_{product} - \sum(\varepsilon_0 + G_{corr})_{reactants}$

Here, $(\varepsilon_0 + G_{corr})_{product}$ are free energies of $H_2$-O-GQD/$H_2$-BC$_2$O-GQD/$H_2$-BCO$_2$-GQD systems, while $(\varepsilon_0 + G_{corr})_{reactants}$ denotes the sum of the free energies of $H_2$ and O-GQD/ BC$_2$O-GQD/ BCO$_2$-GQD, respectively. The above Σ denotes the sum over different parts of the system (reactant or product), while $\varepsilon_0$ and $G_{corr}$, respectively, denote the total electronic energy and the correction to the Gibbs free energy due to internal energy, for each part [80]. The values of ΔG of $H_2$-O-GQD, $H_2$-BC$_2$O-GQD and $H_2$-BCO$_2$-GQD calculated using this approach are 0.213 eV, 0.183 eV and 0.208, respectively, which are in very good agreement with the values computed using Eq. 5. Figure 6 displays the free-energy diagrams for HER through $H_2$-O-GQD, $H_2$-BC$_2$O-GQD, and $H_2$-BCO$_2$-GQD, at 298.15 K. The HER pathway primarily consists of three steps: (a) initial $H^+ + e^-$, (b) an intermediate adsorbed H*, and (c) final production of $H_2$ molecule [48,81].

**Table 2:** Calculated adsorption energy ($E_{ads}$), Gibbs free energy (ΔG), open-circuit potential V and overpotential (η) of all considered systems.

| System | ΔE$_{ads}$ (eV) | ΔG (eV) | open-circuit potential V (V) | overpotential η (V) |
|---|---|---|---|---|
| $H_2$-O-GQD | -0.059 | 0.181 | -0.33 | 0.181 |
| $H_2$-BC$_2$O-GQD | -0.031 | 0.209 | -1.04 | 0.209 |
| $H_2$-BCO$_2$-GQD | -0.032 | 0.206 | -1.16 | 0.206 |

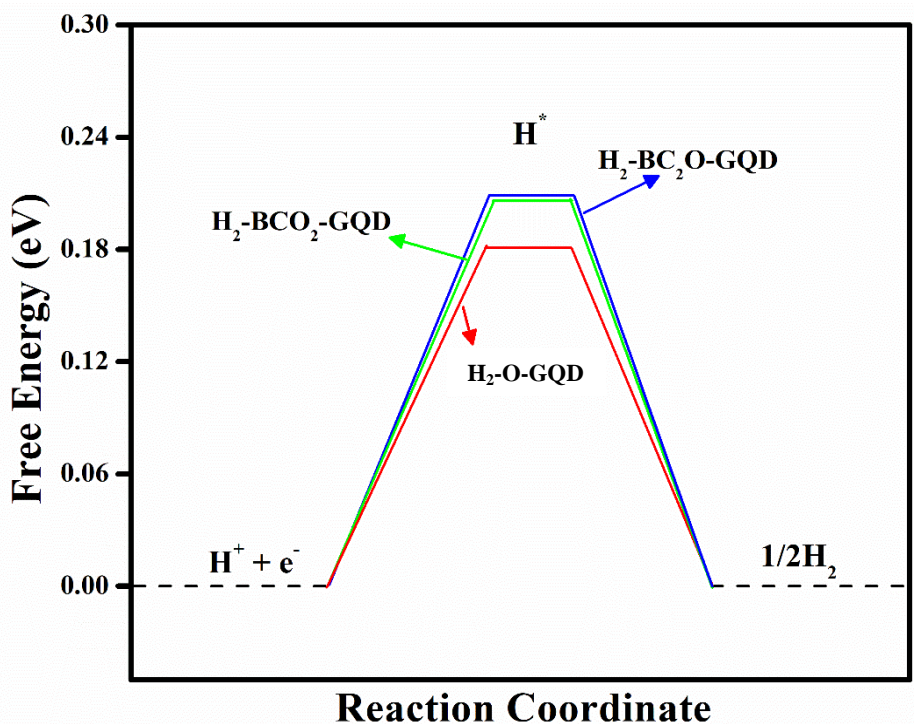

**Figure 6:** Reaction diagram starting from initial $H^+ + e^-$ and via intermediate adsorbed $2H^*$ resulting in generated $H_2$ molecules.

Because, the free-energy of adsorption of 2H* is regarded as a better descriptor of the HER activity than that of 1H* [82-83], therefore, we have calculated that quantity, and presented that in Table 2. We note that ΔG (0.181 eV) of $H_2$-O-GQD is more favorable as compared to $H_2$-$BC_2$O-GQD (0.209 eV) and $H_2$-$BCO_2$-GQD (0.206 eV). This is attributed to its lower adsorption energy in comparison with the other two systems. As mentioned previously, our calculated value of $E_{ads}$ with O-GQD suggests equivalent HER performance as compared to the palladium and platinum surface [84]. Additionally, $H_2$-O-GQD, $H_2$-$BC_2$O-GQD and $H_2$-$BCO_2$-GQD present better HER activity than H monomer over pyrene and coronene polyaromatic hydrocarbons as their binding energy ranges between 0.6–1.6 eV and 0.6–1.4 eV respectively [85].

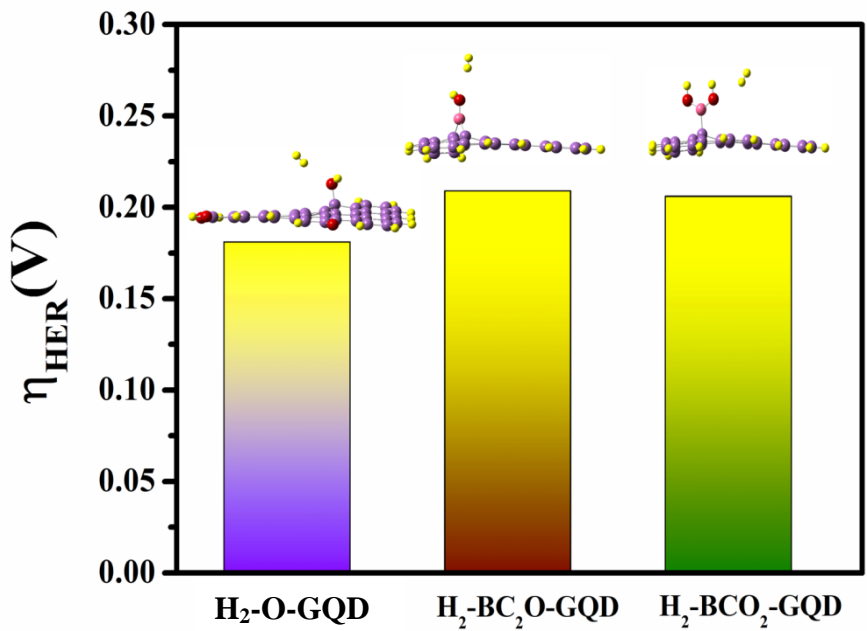

**Figure 7:** Calculated overpotentials for hydrogen over (a) $H_2$-O-GQD, (b) $H_2$-$BC_2$O-GQD and (c) $H_2$-$BCO_2$-GQD.

Using eqn. 6, the overpotential ($\eta$) is evaluated and plotted in Fig. 7 for $H_2$-O-GQD, $H_2$-$BC_2$O-GQD and $H_2$-$BCO_2$-GQD. Overpotential $\eta$ is defined as the minimum potential difference between the two electrodes required for starting the electrochemical water-splitting reaction. The role of electrocatalysts is also to lower this $\eta$ thereby enabling a highly efficient reaction. It is seen from Fig. 7 and Table 2 that $H_2$-O-GQD has the least value of $\eta$ (0.181 V), thus, indicating better catalytic activity as compared to $H_2$-$BC_2$O-GQD and $H_2$-$BCO_2$-GQD. The value of the overpotential for an ideal catalyst for HER is zero, and our calculated values are fairly close to it. Thus, we believe that the structures considered here are strong candidates as catalysts for HER, capable of possibly replacing traditional and costly Pt-based catalysts [86-87].

## Conclusion

In present study, functionalized graphene quantum dots are studied using first principles calculations based on density functional theory. The ground state properties, electronic properties and hydrogen evolution reaction (HER) catalytic activities are examined for oxygen and boron functionalized GQDs (O-GQD, $BC_2O$-GQD and $BCO_2$-GQD). The functionalization significantly alters structural and electronic characteristics of GQDs. Our calculations suggest that $H_2$-O-GQD, with the smallest HOMO-LUMO gap of 1.192 eV compared to other $H_2$ adsorbed GQDs, will have the best electrocatalytic performance in HER. In pursuit of efficient HER catalyst among the considered systems, adsorption and Gibbs free energy is calculated. The calculated $E_{ads}$ of $H_2$-O-GQD present superior energies with -0.059 eV value as compared to $H_2$-$BC_2O$-GQD and $H_2$-$BCO_2$-GQD. Furthermore, as is obvious from the numbers presented in Table 2, the values of other quantities such as adsorption energy, work function, open-circuit potential, overpotential, and the Gibbs free energy of $H_2$-O-GQD are superior as compared to the corresponding values for other GQDs. The Gibbs free energy of $H_2$-O-GQD (0.181 eV) is more favorable as compared to $H_2$-$BC_2O$-GQD (0.209 eV) and $H_2$-$BCO_2$-GQD (0.206 eV). For an ideal catalyst, value of the overpotential for HER is zero, and our calculated values are fairly close to it ranging from 0.18 to 0.2 V. One can always argue that the size of the GQDs considered in this work is not large enough when compared to nanometre sized structures considered in the experiments. In response to that we would like to state that we may not be able to achieve full quantitative agreement with the experiments due to this size difference, but, we strongly believed that our work will provide important qualitative insights and trends which will be useful for the experimentalists. Additionally, we found that the catalytic parameters of $H_2$-O-GQD are quite comparable to those of Pt and Pd surfaces, which, if true, can have very far-reaching implications in the field. Therefore, we urge the experimentalists to investigate the catalytic performance of the structures considered in this work.

# Acknowledgements

Authors VS and BR would like to acknowledge the support by Institute Post-Doctoral Fellowship (IPDF) of the Indian Institute of Technology Bombay.